\documentclass[twocolumn,conference]{IEEEtran}
\usepackage[T1]{fontenc}
\usepackage{units}
\usepackage{amsmath}
\usepackage{amssymb}
\usepackage{graphicx}

\makeatletter
\usepackage[caption=false,font=footnotesize]{subfig}

\makeatother

\begin{document}

\title{Energy Saving of Base Stations Sleep Scheduling for Multi-Hop Vehicular
Networks}

\author{\IEEEauthorblockN{Tao Han\IEEEauthorrefmark{1}, Xiong Liu\IEEEauthorrefmark{1}, Lijun
Wang\IEEEauthorrefmark{2}, Jiang Wang\IEEEauthorrefmark{3}, Kyung
Sup Kwak\IEEEauthorrefmark{4} and Qiang Li\IEEEauthorrefmark{1}}\IEEEauthorblockA{\IEEEauthorrefmark{1}School of Electronic Information and Communications,
Huazhong University of Science and Technology, Wuhan, China}\IEEEauthorblockA{\IEEEauthorrefmark{2}Department of Information Science and Technology,
Wenhua College, Wuhan, China}\IEEEauthorblockA{\IEEEauthorrefmark{3}Shanghai Research Center for Wireless Communication,
Shanghai, China}\IEEEauthorblockA{\IEEEauthorrefmark{3}Shanghai Institute of Microsystem and Information
Technology, Chinese Academy of Sciences, Shanghai, China}\IEEEauthorblockA{\IEEEauthorrefmark{4}Inha Hanlim Fellow Professor, Department of
Information and Communication, Inha university, Incheon, Korea}\IEEEauthorblockA{Email: \IEEEauthorrefmark{1}\{hantao, m201571774, qli\_patrick\}@hust.edu.cn,
\IEEEauthorrefmark{2}wanglj22@163.com, \IEEEauthorrefmark{3}jiang.wang@wico.sh,
\IEEEauthorrefmark{4}kskwak@inha.ac.kr}}
\maketitle
\begin{abstract}
This paper investigates the energy saving of base station (BS) deployed
in a 1-D multi-hop vehicular network with sleep scheduling strategy.
We consider cooperative BS scheduling strategy where BSs can switch
between sleep and active modes to reduce the average energy consumption
utilizing the information of vehicular speeds and locations. Assuming
a Poisson distribution of vehicles, we derive an appropriate probability
distribution function of distance between two adjacent cluster heads,
where a cluster is a maximal set of vehicles in which every two adjacent
vehicles can communicate directly when their Euclidean distance is
less than or equal to a threshold, known as the communication range
of vehicles. Furthermore, the expected value of the sojourn time in
the sleep mode and energy saving are obtained. The numerical results
show that the sleep scheduling strategy significantly reduces the
energy consumption of the base stations.\footnote{The corresponding author is Lijun Wang. The authors would like to
acknowledge the support from the International Science and Technology
Cooperation Program of China (Grant No. 2015DFG12580 and 2014DFA11640),
the National Natural Science Foundation of China (NSFC) (Grant No.
61471180, 61301128, 61461136004, and 61271224), the NSFC Major International
Joint Research Project (Grant No. 61210002), Hubei Provincial Department
of Education Scientific research projects (Grant No. B2015188), Shanghai
Action plan of science and technology innovation (Grant No. 15511103200),
the Fundamental Research Funds for the Central Universities (HUST
Grant No. 2015MS038), the Special Research Fund for the Doctoral Program
of Higher Education (Grant No. 20130142120044), Wenhua College (Grant
No. 2013Y08), and EU FP7-PEOPLE-IRSES (Contract/Grant No. 247083,
318992 and 610524). Kyung Sup Kwak's work is supported by National
Research Foundation of Korea-Grant funded by the Korean Government
(Ministry of Science, ICT and Future Planning-NRF-2014K1A3A1A20034987).} 
\end{abstract}


\section{Introduction}

Vehicular ad hoc network (VANET) is a mobile multi-hop network formed
by vehicles traveling on the road and utilizes multi-hop communication
to improve connectivity probability of every vehicle. In conventional
infrastructure-based vehicular communication networks, information
are transmitted to every vehicle by base stations (BSs) deployed alongside
the road. While in VANETs, two vehicles can communicate with one another
directly as long as the Euclidean distance between two vehicles is
less than a threshold, the communication range of a vehicle, defined
as $r_{0}$ in this paper \cite{7102933_1}. Due to the mobility
of vehicles, the topology of a VANET is changing over time \cite{6814051_2},
so the clusters of vehicles are splitting and merging over time, where
a cluster is a maximal set of vehicles in which there is at least
one hop between each pair of vehicles. Therefore, the information
propagation is distinct from conventional wireless networks.

The ever increasing data demand of vehicles leads to more operation
power of BSs, assumed as the linear function of transmission power,
in infrastructure-based vehicular networks \cite{6963798_3}. Especially
when there is no vehicle on the road, e.g., at night, BS's operation
will waste lots of energy. A useful solution is to turn off the BS
if no vehicle is in its range \cite{7041163_4}. The dynamic network
topology makes it difficult to design a mechanism of BS sleeping schedule
\cite{6416884_5}, especially when multi-hop communication among vehicles
are considered. Multi-hop communication allows a BS only need to communicate
to only one vehicle of a cluster, then by vehicle to vehicle communication,
all vehicles in a cluster can receive or transmit messages.

In this paper, our main contributions are as follows: 1) we develop
a statistic energy saving model for 1-D multi-hop vehicular network
with BSs uniformly deployed at roadside; 2) the probability distribution
function (PDF) for the distance between two adjacent cluster heads
is provided; 3) the expected value of the sojourn time in the sleep
mode and the expected energy saving of a BS are obtained; 4) the impact
of network parameters on energy saving is investigated. 

The rest of this paper is organized as follows: Section II reviews
related work. Section III describes the system model. The analysis
on the energy efficiency is presented in Section IV. Section V discusses
the impacts of network parameters on energy saving. Finally Section
VI concludes this paper and proposes future work.

\section{Related Work}

In recent years, VANETs have attracted significant interest due to
their large number of potential applications. Most of them consider
the connectivity of vehicles or algorithm for cluster formation, the
authors of \cite{6906442_6} took vehicle to infrastructure communication
and vehicle to vehicle communication into consideration, connectivity
probability of platoon-based VANETs where vehicles follow a Poisson
distribution is derived, and they investigated the relationship between
connectivity probability and network parameters. In \cite{5753961_7},
Zhang et al. considered a model in which time is divided into time
slots of equal length and each vehicle changes its speed at the beginning
of each time slot, independent of its speed in other time slots. They
derived a closed form expression for the PDF of cluster length using
queue theory. The authors of \cite{7157957_8} investigated the optimum
deployment of road side units to maximize connectivity of vehicular
networks. As for BS sleeping algorithm, there are a lot of research
in this field. \cite{6364240_9} proposed a switch on/off algorithm
for BSs, which exploits the knowledge of the distance between the
user equipments and their associated BSs in the LTE-Advanced cellular
networks. They obtained the relation between energy consumption and
time of day. In \cite{5723132_10}, Lee et al. proposed an energy
efficient power control mechanism for BS in mobile communication system
in order to improve efficient use of BS transmission power. BSs monitor
the number of nodes in sector and controls transmission power accordingly.
In \cite{7022905_11}, Han et al. investigated the energy efficiency
of BS sleep scheduling strategies in 1-D infrastructure-based vehicular
networks where vehicles are Poissonly distributed and the vehicle
speeds are uniformly distributed. BSs can schedule their sleep by
utilizing the information of vehicles' location and speed. Analytical
results on the expected amount of energy saving of a BSs are derived,
which is shown to be a function of vehicular density and speed distribution.
The study in \cite{7022905_11} did not consider vehicle-to-vehicle
communication which will greatly reduce the energy consumption as
illustrated in this paper afterwards.

\section{System Model}

This paper considers a vehicular network employing both vehicle to
vehicle communication and vehicle to infrastructure communication.
The spatial distribution of vehicles follows a homogeneous Poisson
process with density $\rho$ \cite{7047349_12}. In this paper, the
speed of each vehicle is considered to be independently and identically
distributed following a common uniform distribution \cite{5590335_13}.
Specifically, the PDF of vehicular speed denoted by $V$ is 
\begin{equation}
\mathit{f_{v}(v)}=\begin{cases}
\frac{1}{b-a},~~ & a<v<b\\
0, & \textrm{otherwise}
\end{cases},
\end{equation}
where $b$ and $a$ represent the maximal and minimal speed, respectively,
and we only consider vehicles traveling in one direction. As commonly
done in this field, we assume that vehicular speed remains constant
over time to simplify the analysis.

\begin{figure}[tbh]
\begin{centering}
\includegraphics[scale=0.45]{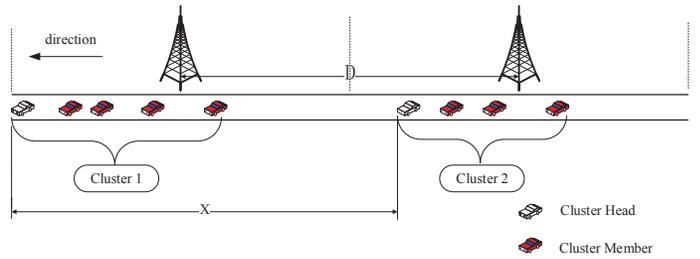}
\par\end{centering}
\caption{\label{fig:1}System Model }
\end{figure}

A set of BSs are regularly deployed with distance $D$ between adjacent
BSs, as illustrated in Fig. \ref{fig:1}. Typical values of $D$ depends
on practical implementations \cite{6503331_14}. The communication
range of a BS is $D$ with BS as the center. 

In this paper, we define the first vehicle in the direction of travel
of a cluster as the cluster head (CH), only the CH can directly communicate
with its nearest BS and other vehicles in the cluster must communicate
with BS via the CH by multi-hop communication.

Further, it is assumed that when a CH arrives or leaves the range
of a BS, the BS can detect it by listening to the indicating signal
sent from the CH. A BS can be in either an active mode or a sleep
mode. In the active mode, a BS operates as normal; while in the sleep
mode, a BS switches off its transmitter and amplifier to save energy.
Therefore, a BS cannot transmit signal but can receive signal only
for detecting CHs in its communication range when it is in the sleep
mode. This paper considers the sleep scheduling strategy that BSs
switch between the active mode and the sleep mode without affecting
the QoS such as connectivity, a BS will keep active as long as there
is a CH in its coverage. 

We next describe the energy consumption model. Let $P_{\mathrm{a}}$
be the power consumed in the active mode which includes air conditioning,
signal processing, power supply and power amplifier, etc. and let
$P_{\mathrm{s}}$ be the power of signal processing unit for transmitting
pilot signals when a BS in a sleep mode. \cite{5956433_15}. So when
a BS is in the sleep mode, the power saved is $P_{0}=P_{\mathrm{a}}-P_{\mathrm{s}}$.
Further, consider that the energy cost of turning a BS off and then
back on is $E_{\mathrm{c1}}$ and $E_{\mathrm{c2}}$ respectively.
Then the total energy cost of turning a BS off and later back on is
$E_{\mathrm{c}}$ = $E_{\mathrm{c1}}$ + $E_{\mathrm{c2}}.$ This
extra energy cost can have a impact on the energy saving achieved
by a BS.

\section{Analysis of Energy Saving}

In this paper, we assume that vehicles are distributed following a
Poisson distribution. Denote by $Y$ the Euclidean distance between
two randomly chosen adjacent vehicles and $X$ the Euclidean distance
between two randomly chosen adjacent CHs at a arbitrary time instant.
It follows from the Poisson distribution of vehicle that the variable
$Y$ follows an exponential distribution with parameter $\rho$. We
define a variable $T$ as $T=\frac{X-D}{V}$ and $f_{t}\mathfrak{\left(\mathit{t}\right)}$
is its PDF. When $X$ is larger than $D$, $T$ represents the time
period from the first cluster head leaves the communication range
of a BS to the next cluster head enters the communication range of
the BS. Then the BS can detect these events so that its transmitter
and amplifier can be switched off to save energy and switched on afterwards,
$T$ is the time period of BS in sleep mode. So the PDF of variable
$X$ is the key point to work out the energy saved by our model.

\subsection{Distribution of the distance between adjacent CHs}

According to \cite{5753961_7}, the PDF of the cluster length $x_{0}$
is 
\begin{eqnarray}
f_{x_{0}}(x_{0}) & = & \frac{\rho}{e^{\rho r_{0}}-1}\overset{\left\lfloor \nicefrac{x_{0}}{r_{0}}\right\rfloor }{\underset{m=0}{\sum}}\frac{\left(-\rho\left(x_{0}-mr_{0}\right)\right)^{m-1}}{-m!}\nonumber \\
 &  & \times\left(\rho(x_{0}-mr_{0})+m\right)e^{-\rho mr_{0}},\label{eq:2}
\end{eqnarray}
the CDF of the distance between two adjacent CHs $X$ is the sum of
cluster length $x_{0}$ and a random truncated exponential variable
$x_{1}$ , the distance between the last vehicle of the cluster and
the CH of the next cluster, whose PDF is 
\begin{equation}
f_{x_{1}}(x_{1})=\begin{cases}
\frac{\rho e^{-\rho x_{1}}}{e^{-\rho r_{0}}},~~ & x_{1}>r_{0}\\
0, & \textrm{otherwise}
\end{cases},
\end{equation}
so the CDF of $X$ can be calculated as 
\begin{eqnarray}
F_{x}(x) & = & \int_{0}^{x-r_{0}}\int_{r_{0}}^{x-x_{0}}f_{x_{0}}(x_{0})\cdot f_{x_{1}}(x_{1})\mathrm{d}x_{1}\mathrm{d}x_{0}\nonumber \\
 & = & \int_{0}^{x-r_{0}}f_{x_{0}}(x_{0})\mathrm{d}x_{0}\int_{r_{0}}^{x-x_{0}}\frac{\rho e^{-\rho x_{1}}}{e^{-\rho r_{0}}}\mathrm{d}x_{1}\nonumber \\
 & = & \int_{0}^{x-r_{0}}f_{x_{0}}(x_{0})(1-e^{-\rho(x-x_{0}-r_{0})})\mathrm{d}x_{0}.
\end{eqnarray}

Therefore, the PDF of $X$ follows that
\begin{eqnarray}
f_{x}(x) & = & \left[\int_{0}^{x-r_{0}}f_{x_{0}}(x_{0})(1-e^{-\rho(x-x_{0}-r_{0})})\mathrm{d}x_{0}\right]^{'}\nonumber \\
 & = & \left[\int_{0}^{x-r_{0}}f_{x_{0}}(x_{0})-e^{-\rho(x-r_{0})}e^{\rho x_{0}}f_{x_{0}}(x_{0})\mathrm{d}x_{0}\right]^{'}\nonumber \\
 & = & \rho e^{-\rho(x-r_{0})}\int_{0}^{x-r_{0}}e^{\rho x_{0}}f_{x_{0}}(x_{0})\mathrm{d}x_{0}.\label{eq:5}
\end{eqnarray}

Substitute the preceding equation \eqref{eq:2} into \eqref{eq:5},
because $f_{x_{0}}(x_{0})$ is a piece-wise function, to calculate
the integral, we need to divide the $x_{0}$ into segments of equal
length $r_{0}$, and define $y=x_{0}-mr_{0}$, when $x\geqslant2r_{0}$,
then we can obtain 
\begin{eqnarray}
f_{x}(x) & = & \frac{-\rho e^{-\rho(x-r_{0})}}{e^{\rho r_{0}}-1}\{\overset{\left\lfloor \nicefrac{x}{r_{0}}-1\right\rfloor }{\underset{k=0}{\sum}}\overset{k-1}{\underset{m=0}{\sum}}\nonumber \\
 &  & \int_{\rho(k-1-m)r_{0}}^{-\rho(k-m)r_{0}}\frac{y^{m}}{m!}e^{-y}-\frac{y^{m-1}}{(m-1)!}e^{-y}\mathrm{d}y\nonumber \\
 &  & +\int_{-\rho(\left\lfloor \nicefrac{x}{r_{0}}-1\right\rfloor -m)r_{0}}^{-\rho(x-mr_{0}-r_{0})}\frac{y^{m}}{m!}e^{-y}-\frac{y^{m-1}}{(m-1)!}e^{-y}\mathrm{d}y\}\nonumber \\
 & = & \frac{\rho e^{-\rho(x-r_{0})}}{e^{\rho r_{0}}-1}\overset{\left\lfloor \nicefrac{x}{r_{0}}-1\right\rfloor }{\underset{k=0}{\sum}}\{\overset{k-1}{\underset{m=0}{\sum}}\nonumber \\
 &  & \frac{e^{\rho(k-m)r_{0}}[-\rho(k-m)r_{0}]^{m}}{m!}\nonumber \\
 &  & -\frac{e^{\rho(k-m-1)r_{0}}[-\rho(k-m-1)r_{0}]^{m}}{m!}\}\nonumber \\
 &  & +\{\frac{e^{\rho(x-kr_{0}-r_{0})}\left[-\rho(x-kr_{0}-r_{0})\right]^{k}}{k!}\nonumber \\
 &  & -\frac{e^{\rho(\left\lfloor \nicefrac{x}{r_{0}}-1\right\rfloor -kr_{0})}\left[-\rho(\left\lfloor \nicefrac{x}{r_{0}}-1\right\rfloor -kr_{0})\right]^{k}}{k!}\},
\end{eqnarray}
when $r_{0}\leqslant x<2r_{0}$, 
\begin{equation}
f_{x}(x)=\frac{\rho[1-e^{-\rho(x-r_{0})}]}{e^{\rho r_{0}}-1}.
\end{equation}

According to the definition of mathematics expectation , we have
\begin{equation}
\mathbb{E}[X]=\int_{r_{0}}^{\infty}xf_{x}(x)\mathrm{d}x.
\end{equation}

\subsection{Time interval of sleep mode}

As mentioned above, CHs transmit signals to BS if they come in or
step off the coverage of the BS, so when a CH leaves the coverage
of a BS and the next CH is not in its coverage, BS can turn off and
then keep in sleep mode until the next CH comes in its coverage.

Thus the time interval of sleep mode for a typical BS can be defined
as

\begin{equation}
T_{\mathrm{off}}=\begin{cases}
T,~~ & T>0\\
0, & T\leq0
\end{cases}.\label{eq:9}
\end{equation}

Apparently, the PDF of $T_{\mathrm{off}}$ is hard to obtain, so we
may pay attention to its expected value which can be derived as 
\begin{eqnarray}
\mathbb{E}[T_{\mathrm{off}}] & = & \frac{\int_{0}^{\infty}tf_{t}(t)\mathrm{d}t}{\mathbb{P}\{T>0\}}\nonumber \\
 & = & \frac{\int_{D}^{\infty}\int_{a}^{b}\frac{x-D}{v}\cdot f_{x}(x)\cdot f_{v}(v)\mathrm{d}v\mathrm{d}x}{\mathbb{P}\{X>D\}}\nonumber \\
 & = & \frac{\ln b-\ln a}{b-a}\cdot\frac{\int_{D}^{\infty}(x-D)f_{x}(x)\mathrm{d}x}{\int_{D}^{\infty}f_{x}(x)\mathrm{d}x}.
\end{eqnarray}

In the following section, we will investigate the relation between
$T_{\mathrm{off}}$ and the system parameters.

\subsection{Average power saved }

In the same way, the active time period of a BS is denoted as $T_{\mathrm{on}}$,
which is 
\begin{equation}
T_{\mathrm{on}}=\begin{cases}
\frac{D}{V},~~ & X>D\\
\frac{X}{V}, & X\leq D
\end{cases}.\label{eq:11}
\end{equation}

The energy saved $E_{\mathrm{\mathrm{off}}}$ in sleep mode can be
valued by
\begin{equation}
E_{\mathrm{\mathrm{off}}}=T_{\mathrm{\mathrm{off}}}P_{0}-E_{\mathrm{c}}.
\end{equation}

We denote $P_{\mathrm{save}}$ as the power saved by our sleeping
scheduling which can be calculated as
\begin{equation}
P_{\mathrm{save}}=\frac{T_{\mathrm{off}}P_{0}-E_{\mathrm{\mathrm{c}}}}{T_{\mathrm{off}}+T_{\mathrm{on}}},\label{eq:13}
\end{equation}
substitute the preceding equation \eqref{eq:9} and \eqref{eq:11}
into \eqref{eq:13}, we get 
\begin{equation}
P_{\mathrm{save}}=\begin{cases}
\frac{(\frac{X-D}{V})P_{0}-E_{\mathrm{c}}}{\nicefrac{X}{V}},~~ & X>D\\
0, & X\leq D
\end{cases}.
\end{equation}

Then the expected value of the power saved for a typical BS denoted
as $\mathbb{E}[P_{\mathrm{save}}]$ can be calculated by
\begin{eqnarray}
\mathbb{E}[P_{\mathrm{save}}] & = & \mathbb{E}[\frac{\frac{X-D}{V}\cdot P_{0}-E_{\mathrm{c}}}{\nicefrac{X}{V}}]\cdot\mathbb{P}\{X>D\}\nonumber \\
 & = & \mathbb{E}[\frac{(X-D)P_{0}-E_{\mathrm{c}}.V}{X}]\cdot\mathbb{P}\{X>D\}\nonumber \\
 & = & \mathbb{E}[P_{0}-\frac{DP_{0}}{X}-E_{\mathrm{c}}\frac{V}{X}]\cdot\mathbb{P}\{X>D\}.
\end{eqnarray}

Due to the independence of $X$ and $D$, we have
\begin{equation}
\mathbb{E}[\frac{V}{X}]=\mathbb{E}[\frac{1}{X}]\cdot\mathbb{E}[V],
\end{equation}
where
\begin{equation}
\mathbb{E}[\frac{1}{X}]=\frac{\int_{D}^{\infty}\frac{1}{x}f_{x}(x)\mathrm{d}x}{\mathbb{P}\{X>D\}},
\end{equation}
\begin{equation}
\mathbb{E}[V]=\frac{a+b}{2},
\end{equation}
and
\begin{equation}
\mathbb{P}\{X>D\}=\int_{D}^{\infty}f_{x}(x)\mathrm{d}x=1-F(D).
\end{equation}

Through the given results, we can obtain that
\begin{eqnarray}
\mathbb{E}[P_{\mathrm{save}}] & = & P_{0}\cdot[1-F(D)]-P_{0}\cdot D\cdot\int_{D}^{\infty}\frac{1}{x}\cdot f_{x}(x)\mathrm{d}x\nonumber \\
 &  & -E_{\mathrm{c}}\cdot\mathbb{E}[V]\cdot\int_{D}^{\infty}\frac{1}{x}\cdot f_{x}(x)\mathrm{d}x.\label{eq:18}
\end{eqnarray}

According to \eqref{eq:18}, we know the $\mathbb{E}[P_{\mathrm{save}}]$
is related to the vehicular average speed and communication range
$r_{0}$. The next section will show the relationship between the
energy saving and these system parameters.

\section{Numerical Results}

This section shows numerical evaluation of the results presented in
the previous sections, followed by discussions and conclusions. The
typical values of system parameters are selected from certain practical
scenarios. Specifically, the distance between BSs is $D=800\,\mathrm{m}$.
The energy consumption model uses parameters $P_{\mathrm{0}}=1\,\mathrm{kw}$
and $E_{\mathrm{c}}=10\,\mathrm{J}$. 

\begin{figure}[tbh]
\begin{centering}
\includegraphics[scale=0.55]{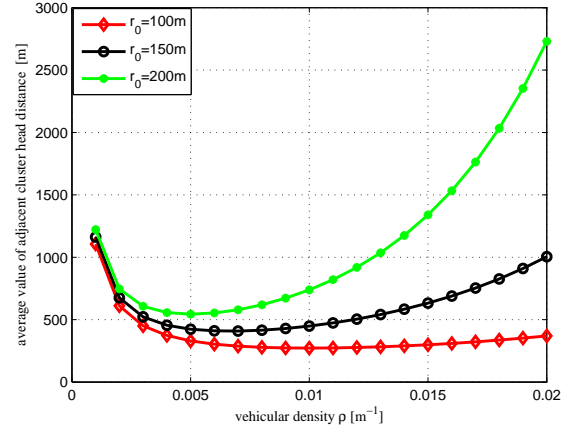}
\par\end{centering}
\caption{\label{fig:2}The expected value of distance between two adjacent
CHs}
\end{figure}

First, the expected value of the distance between two adjacent CHs
shown in Fig. \ref{fig:2} is related to vehicular density and communication
range. Undoubtedly, the larger $r_{0}$ means that vehicles have more
opportunity to directly communicate with adjacent vehicles, so leading
to more vehicles in a cluster and then larger value of $\mathbb{E}[X]$.
And for a certain $r_{0}$, the value of $\mathbb{E}[X]$ decreases
first and then increases to infinite as the vehicular density grows.
In the case of very low vehicular density, the distance between two
adjacent vehicles seems to be large enough and they cannot communicate
directly, then the distance of two adjacent CHs is very large. With
the increase of vehicular density, the distance between two adjacent
vehicles decrease and some of them can communicate directly, so the
value of $\mathbb{E}[X]$ may decrease, but if the vehicular density
continuously increase, the distance of two adjacent vehicles all become
less than $r_{0}$, causing a large amount of vehicles in a cluster,
leading to the increase of the value of $\mathbb{E}[X]$.

\begin{figure}[tbh]
\begin{centering}
\includegraphics[scale=0.55]{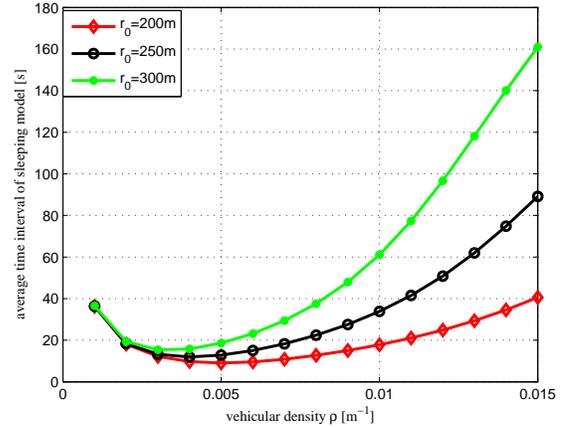}
\par\end{centering}
\caption{\label{fig:3}Time interval of sleep mode versus vehicular transmission
range, where $a=40\,\mathrm{km/h}$, $b=80\,\mathrm{km/h}$.}
\end{figure}

Then we investigate the relationship between the sojourn time of sleep
mode and traffic parameters, such as vehicular density $\rho$ and
transmission range $r_{0}$. Since the sojourn time in sleep model
is derived from the $\mathbb{E}[X]$, the trend of $\mathbb{E}[T_{\mathrm{off}}]$
is the same with $\mathbb{E}[X]$, which is clearly depicted in Fig.
\ref{fig:3}. 

\begin{figure}[tbh]
\begin{centering}
\includegraphics[scale=0.55]{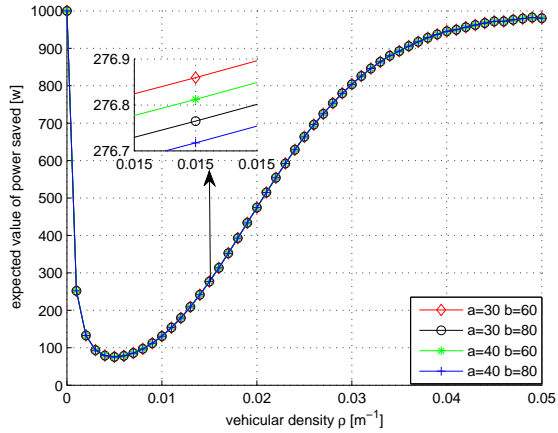}
\par\end{centering}
\caption{\label{fig:4}Average power saved, $r_{0}=200\,\mathrm{m}$. }
\end{figure}

Next we come to the average power saved $\mathbb{E}[P_{\mathrm{save}}]$,
the relationship between $\mathbb{E}[P_{\mathrm{save}}]$ and vehicular
density is clearly presented in Fig. \ref{fig:4}. The larger value
of $\mathbb{E}[X]$ means the more time for BS keeping in sleep model,
so the curve of $\mathbb{E}[P_{\mathrm{save}}]$ is similar with $\mathbb{E}[X]$.
The average value of vehicular speed together with the energy consumption
of switching BS between active and sleep mode will affect the average
power saved, as reflected in the third part of \eqref{eq:18}. When
the speed of vehicles is higher, the sojourn time in sleep model is
less, so leading to less energy saving. Because of the small value
of $\int_{D}^{\infty}\frac{1}{x}f_{x}(x)\mathrm{d}x$, the difference
between different distributions is nearly negligible as depicted in
Fig. \ref{fig:4}. While considering a ideal condition that the handover
cost is zero, the speed distribution will no longer influence the
energy saving. 

\begin{figure}[tbh]
\begin{centering}
\includegraphics[scale=0.55]{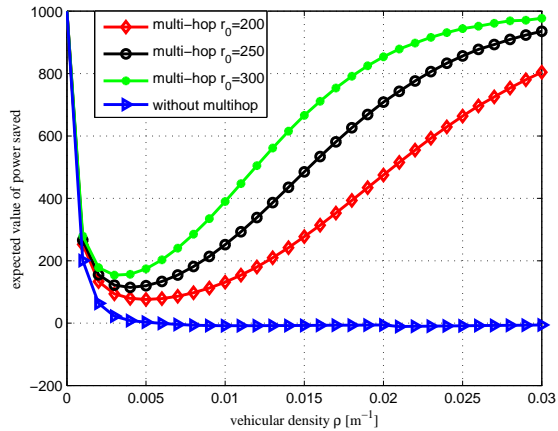}
\par\end{centering}
\caption{\label{fig:Our-sleeping-strategy}Our sleeping strategy compares with
the scenario without multi-hop communication, where $a=40\,\mathrm{km/h}$,
$b=80\,\mathrm{km/h}$.}
\end{figure}

Since our proposed sleeping schedule takes multi-hop communication
into consideration, we wold like to compare our analytical results
with the scenario proposed in \cite{7022905_11}. As we can see in
Fig. \ref{fig:Our-sleeping-strategy}, our strategy has a better performance
especially when the vehicular density is high, in this situation.
we can take full advantage of the direct communication between vehicles
to save energy of BSs, while in the situation without multi-hop, BSs
try to communicate with every vehicles in its coverage, causing a
huge amount of energy consumption. 

\section{Conclusions and Future Work}

This paper considers a multi-hop communication via connections among
vehicles, and analyzes the energy efficiency of BS sleeping strategy
in 1-D cluster-based vehicular networks. Analytical results have been
provided for the expected time interval in sleep mode of a BS. We
also investigate the effects caused by network parameters such as
vehicular speed distribution. Comparing with the sleeping strategy
without considering the multi-hop communication between vehicles,
our sleeping schedule has a great improvement in energy saving of
BSs.

In our system model, a cluster is formed only by vehicles whose distance
with adjacent vehicles is less than $r_{0}$, so the clusters are
not stable for vehicular speed changes over time therefore intervals
also change, this is not beneficial for information transmission.
Then in the future work, we may propose a better ways to form clusters
and analyze the energy efficiency, in which two adjacent vehicles
in a cluster satisfy that the distance between them is less than a
threshold, and their speeds are close enough as well.

\bibliographystyle{IEEEtran}
\bibliography{VANET_energy_saving}

\end{document}